# Ab initio calculated dynamic structure factors and optical properties of beryllium along the Hugoniot

Wei-Jie Li[1,3*], Jie Zhou[1,3], Zi Li[2,5], Yunliang Zhu[1,3], Han-Dong Hu[1,3,7], Hao Ma[4], Zhe Ma[1,3*], Cong Wang[2,5,6*], Ping Zhang[2,5,6]

[1] Intelligent Science & Technology Academy Limited of CASIC, Beijing, 100141, People's Republic of China

[2] Institute of Applied Physics and Computational Mathematics, Beijing, 100088, People's Republic of China

[3] Scientific Research Key Laboratory of Aerospace Defence Intelligent Systems and Technology, Beijing, 100141, People's Republic of China

[4] School of Mechanical and Electric Engineering, Sanming University, Sanming, 365004, Fujian Province, China
[5] Tianfu Innovation Energy Establishment, Chengdu, 610213, China
[6] Center for Applied Physics and Technology, Peking University, Beijing, 100871, People's Republic of China
[7] Shanghai Jiao Tong University, 200240, People's Republic of China

**Abstract**
Beryllium is an ablator material in the inertial-confinement fusion and hypervelocity impact studies. The thermoelastic properties, structure factors, and optical properties of beryllium are important in these studies. In this paper, the static structure factors, ion-ion dynamic structure factors, adiabatic velocity, and optical properties of beryllium along the Hugoniot were calculated by ab initio simulations. The static structure factors show that beryllium atoms were randomly distributed. The dynamic structure factors via the scattering function have extracted the dispersion relation for the collective excitations. By collecting the peak position of dynamic structure factors, the dispersion relation and adiabatic sound velocity were derived by definitions. By the calculated equation of state, the thermoelastic properties and adiabatic sound velocity have been derived. The two calculated methods about adiabatic sound velocity were verified to be equivalent.

## 1. Introduction

Beryllium is an ablator material in inertial-confinement fusion and hypervelocity impact studies. The physical properties, such as the equations of states (EOS), thermoelastic properties, structure factors, and optical properties are of fundamental importance in understanding the physical process and material composition identification. The EOS are usually determined by shocked experiments or ab initio calculations. The static structure factors $S(q)$ can be measured directly by diffraction experiments. The experimental methods for dynamic structure factors are based on scattering experiments using either fast electrons or x-rays. On the other hand, the dynamic structure factors are closely connected with the dielectric function via the fluctuation-dissipation theorem and all other quantities can be derived. The structure factors can be calculated by the trace of equilibrium structure by its definition. Dynamic ion structure factors of dense plasmas and the warm dense matter is of great importance as it contains the complete information on the ions in these strongly interacting systems and is also influenced by the electron properties.

Experimentally, the principal Hugoniot, sound velocity, and shear stress of beryllium were measured using magnetically accelerated fly plates on the Sandia Z machine [1]. The dynamics structure factors are the Fourier transform of space- and time-dependent density-density correlation function, which furnished a direct measure of the density-fluctuation spectrum of the electrons of scatter. The dynamic structure factors of beryllium was determined by inelastic x-ray scattering experiments [2]. The multiphase EOS were constructed by functional theory calculation, and thermoelastic data was by experimental and resonant ultrasound spectroscopy measurements [3]. Sound velocity can be obtained from inelastic x-ray scattering experiment [4] or theoretical methods [5].

The EOS and ion transport properties of warm dense beryllium were calculated by quantum molecular dynamics simulations [6]. The expanded beryllium evolution from warm dense matter to atomic fluid were calculated by molecular dynamics simulations [7]. The electrical and optical properties of beryllium along the principal Hugoniot were investigated by using quantum molecular dynamics simulations combined with the Kubo-Greenwood formulation [8]. The electron heat capacity and electron-phonon coupling factor of beryllium were calculated ab initio theoretical studies [9]. The premelting hcp to bcc transition of beryllium was calculated by phonon quasiparticles with anharmonic effects taking into consideration [10]. Then, the high-pressure phase diagram of beryllium was calculated by first-principles molecular dynamics coupled with the thermodynamic integration method [11]. Moreover, the dielectric function by the Kubo-Green formula was used to determine the dynamic structure factors [12]. The x-ray Thomson scattering, ion-ion structure factors, compressibilities, phonons, and conductivity of beryllium by supposed density functional neutral-pseudoatom hypernetted-chain model, which agrees closely with density-functional-theory molecular-dynamics simulations [13]. In conclusion, the ab initio simulation can collect the structure factors, EOS, thermoelastic properties, sound velocity and optical properties concurrently.

The comprehensive data about the ion-ion dynamic structure factors allows for the calculation of the dispersion relation for collective excitations. On the one hand, the static and ion-ion dynamic structure factors can be derived by the ab initio calculated equilibrium configurations, and the adiabatic sound velocity by the dispersion relationship derived from the ion-ion dynamics structure factors. On the other hand, the dynamic structure factors are the spectral function of the density-density correlations in the systems and is needed to determine the EOS and transport properties. After a series of ab initio simulations, the EOS of beryllium is collected, which can derive the thermoelastic properties by its definition, such as isothermal bulk modulus and adiabatic sound velocity. From the physical significance, the two methods of adiabatic sound velocity are equivalent. In this paper, the ion-ion dynamic structure factors and optical properties of beryllium along the principal Hugoniot curve were calculated by the ab initio molecular dynamics. Then, the adiabatic sound velocities were derived by both the dispersion relationship and the EOS, which check the equivalence of these two methods.

## 2. Methods and calculation

### 2.1. Structure factor
#### 2.1.1. Static structure factor
The radial distribution function $g(r)$ is defined as

$$g(r) = \frac{V}{4\pi r^2 N^2} \left\langle \sum_{i=1}^{N} \sum_{j=1, j \neq i}^{N} \delta(r - |\mathbf{r}_i - \mathbf{r}_j|) \right\rangle \tag{1}$$

where $V$ is the cell volume, $N$ is the number of atoms, $r_i$ and $r_j$ are atomic coordinates of atoms $i$ and $j$, and $\langle \ldots \rangle$ means the time or ensemble average.

The definition of static structure factors $S(q)$ is

$$S(q) = \frac{1}{N} \left\langle \sum_{i=1}^{N} \sum_{j=1}^{N} e^{i\mathbf{q} \cdot (\mathbf{r}_i - \mathbf{r}_j)} \right\rangle \tag{2}$$

where the wave vector $q$ is determined by $2\pi\sqrt{n}/a$ in which a is the lattice parameters of the cell and $n=0,1,2\ldots$ . To be pointed out, $S(q)$ is computed via the Fourier transform of $g(r)$. The $S(q)$ at small $q$ represents the long-ranged structural information of systems. The isothermal compressibility $\chi$ ($1/K_T$, $K_T$ is isothermal bulk modulus) can be computed from $S(q)$ at $q\to 0$ with the form[14] of

$$\chi = \frac{\lim_{q \to 0} S(q)}{n_i k_B T} = \frac{1}{K_T} \tag{3}$$

where $n_i = N/V$ is the ionic density, $k_B$ is the Boltzmann constant and $T$ is the temperature. The cubic fitting to yield $S(0)$ is enough.

#### 2.1.2. Dynamic structure factor
From the theory of simple liquid [14], the ion-ion dynamic structure factors $S(q,\omega)$ of warm dense matter in thermodynamic equilibrium[15] characterizes the collective dynamics of fluctuations of ionic density over both length and time scales, where $\omega$ denotes the frequency. For a system in thermodynamic equilibrium, the $S(q,\omega)$ is computed via the Fourier transform of intermediate scattering function $F(q,t)$ [16]

$$S(\mathbf{q},\omega) = \frac{1}{2\pi} \int_{-\infty}^{+\infty} F(\mathbf{q},t) e^{i\omega t} dt \tag{4}$$

The $F(q,t)$ is defined as

$$F(q,t) = \frac{1}{N} \langle \rho(\mathbf{q},t) \rho(-\mathbf{q},t) \rangle \tag{5}$$

where $\rho(\mathbf{q},t)$ is the Fourier transform of the real space time-dependent density distribution and contains the ion information at wave number $\mathbf{q}$ and time t

$$\rho(\mathbf{k},t) = \sum_{j=0}^{N} \exp[i\mathbf{k} \cdot \mathbf{r}_j(t)] \tag{6}$$

By collecting the peak position of $S(q,\omega)$ at different wave numbers $q$, the dispersion relation is obtained. The adiabatic speed of sound is estimated by $d\omega/dq|_{q\to 0}$ [15].

2.2. Thermoelastic properties and adiabatic sound velocity by the equation of state

On the other hand, the adiabatic sound velocity can be calculated from the ab initio calculated EOS. From the EOS, the thermoelastic properties can be calculated directly by its definition, such as isothermal bulk modulus, thermal expansion coefficient, heat capacity, and Grüneisen parameter [17].

The isothermal bulk modulus ($K_T$) is

$$K_T = -V\left(\frac{\partial P}{\partial V}\right)_T \tag{7}$$

The thermal expansion coefficient ($\alpha$) is

$$\alpha = \frac{1}{V}\left(\frac{\partial V}{\partial T}\right)_P \tag{8}$$

The heat capacity at a constant volume ($C_V$) is

$$C_V = \left(\frac{\partial E}{\partial T}\right)_V \tag{9}$$

The Grüneisen parameter ($\gamma$) is

$$\gamma = V\left(\frac{\partial P}{\partial E}\right)_V = \frac{\alpha K_T V}{C_V} \tag{10}$$

The adiabatic bulk modulus ($K_s$) is

$$K_s = K_T\left[1 + \alpha\gamma T\right] \tag{11}$$

The adiabatic sound velocity ($V_P$) is

$$V_P = \sqrt{\frac{K_s}{\rho}} \tag{12}$$

Then, the adiabatic sound velocity at specific temperature and pressure states $V_P(T,P)$ are collected by the calculated isothermal bulk modulus, thermal expansion coefficient, and Grüneisen parameter at the same state.

2.3. Optical property

In current ab initio simulations, the optical properties are calculated by the dielectric function. From the Kubo-Greenwood formula, the frequency dependent dielectric function [18] $\varepsilon(\omega) = \varepsilon_1(\omega) + i\varepsilon_2(\omega)$ is collected. The $\varepsilon_1(\omega)$ and $\varepsilon_2(\omega)$ are the real and imaginary parts of the dynamical dielectric response function $\varepsilon(\omega)$. Then, the main optical spectra, such as the reflectivity R(ω), adsorption coefficient I(ω), energy-loss spectrum L(ω), and refractive index n(ω) can be obtained from the dynamical dielectric response function $\varepsilon(\omega)$.

Then the absorption coefficient $I(\omega)$ is

$$I(\omega) = \sqrt{2}\omega\left[\sqrt{\varepsilon_1(\omega)^2 + \varepsilon_2(\omega)^2} - \varepsilon_1(\omega)\right]^{1/2} \quad (13)$$

The energy-loss spectrum $L(\omega)$ is

$$L(\omega) = \varepsilon_2(\omega) / \left[\varepsilon_1(\omega)^2 + \varepsilon_2(\omega)^2\right] \quad (14)$$

The absorption coefficient and emission coefficient are either continuous or linear. The simulating optical spectra of atomic and molecular gases at different thermodynamic conditions are calculated based on the NIST databases and semi-classical quantum theory, which correspond to the line-by-line spectrum. For the continuous spectrum, the black body radiation theory is usually adopted for the solid or liquid. The difference between the black body radiation theory, semi-classical quantum theory, and our calculated optical properties are discussed in this study.

### 2.4. Calculation details

The *ab initio* molecular dynamics calculations are implemented in the plane wave density functional VASP code [19,20]. The projector augmented waves (PAWs) [21,22] and generalized gradient approximation (GGA) in the parameterization of Perdew, Burke, and Ernzerhof [23] were adopted. The plane wave cutoff was 800 eV, which was sufficient to ensure that the pressure converges with 1% accuracy. The selected time step was 1 fs in all the calculations. An accurate calculation of the dynamic structure factors requires ensemble averaged over a large number of particles and a long time scale. We selected 216 atoms as the cell and beryllium atoms were randomly distributed in the cell. The density and temperature at Hugoniot state were referred from Ref. [6,8]. The density ranged from 3.8 g/cm$^3$ to 5.8 g/cm$^3$, and the temperatures ranged from 0.95 eV to 10.65 eV. The long-term correlation function was collected by the *NVT* ensemble with the equilibrium volume, and when the total time exceeded 20 ps. For the calculation of ion-ion dynamic structure factors, the ion movement was extracted from the trajectories of the equilibrium configurations. Then, the optical property was calculated using the Kubo-Greenwood formula and the Chester-Thellung-Kubo-Greenwood formula as implemented in VASP. The optical property was obtained by averaging ten snapshots extracted from the last 6,000 time steps in each MD trajectory with an interval of 600 time steps.

### 3. Results

### 3.1. Static structure factor

The static structure factors of beryllium were calculated by Eq. (2) and shown in Figure 1. An ab initio simulation for these conditions has already been carried out [6,8]. Our simulation uses a larger energy cutoff, a larger number of ions, and a larger number of time steps, which likely leads to more accurate results. The shapes of the static structure factors at high temperatures were consistent with the reported experimental results [13]. The static structure factors showed that beryllium was randomly distributed in the cell, which was consistent with the analysis of the radial distribution function. As density increased, the main peak position of S(q) increased and the peak strength decreased, shown in the subfigure of Figure 1. At higher pressure along the principal Hugoniot state, the mean atom distance decreased as the density increases, and the system was distributed homogeneously. In the long-wavelength limit, $\lim_{q \to 0} S(q)$ was correlated the isothermal compressibility by Eq. (3), and the isothermal compressibility was determined from the EOS by Eq. (7). The $\lim_{q \to 0} S(q)$ increased as pressure increases, which affected the adiabatic sound velocity. The calculated long-wavelength limit $\lim_{q \to 0} S(q) = 0.046$ for T=0.95 eV and ρ=3.8 g/cm³ was consistent with the $\lim_{q \to 0} S(q) = 0.046$ by isothermal compressibility from the early reported EOS [6] at the same state.

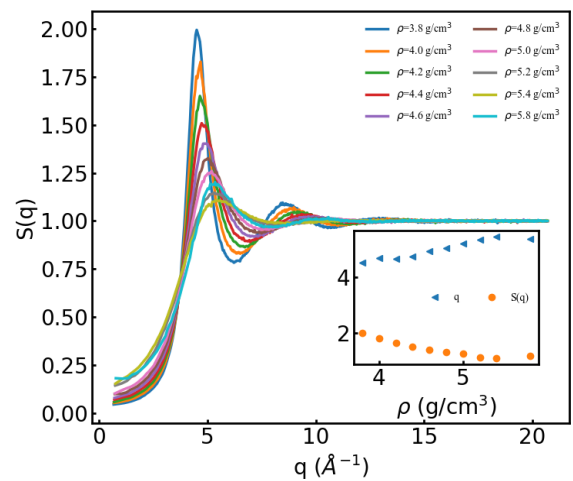

Figure 1 Static structure factors of Be along the Hugoniot labeled by density. The subfigure is the maximum peak of S(q) with different q and S(q).

### 3.2. Dynamic structure factors

The dynamic structure factors S(q,ω) via the intermediate scattering function was collected from Eq. (4), shown in Figure 2(a). It exhibited the well-known sharp plasmon peak below a critical wave vector $q_c$ [24]. The small peak was attributed to the limited number of atoms in the cell with only 256 atoms. The dispersion relation of the collective excitations has been determined by analyzing the position of the side peaks. The position of the peak of the resonance against the wave number yielded the dispersion relation for a different state, shown in Figure 2(b). The slope of the dispersion relation for the small wave vector was the adiabatic sound velocity, and the corresponding sound velocities were determined via linear fits to the dispersion relation at small wave vectors $d\omega/dq|_{q\to 0}$. By fitting the dispersion relation, the adiabatic sound velocity was 16.22 km/s at ρ=3.8 g/cm³ and 20.77 km/s at ρ=5.2 g/cm³.

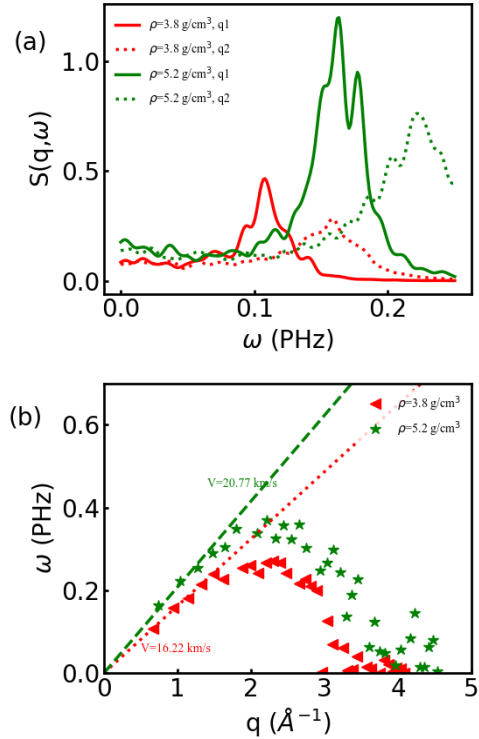

Figure 2 The ab initio calculated (a) dynamic structure factors and (b) dispersion curve.

3.3. Adiabatic sound velocity by dynamic structure factors and EOS

From the early reported EOS of beryllium [6], the thermoelastic properties along the Hugoniot state were collected. The isothermal compressibility (bulk modulus) calculated by EOS was directly used in the calculation of the adiabatic sound velocity. The adiabatic sound velocities of beryllium along the Hugoniot state calculated by EOS and dynamic structure factors were shown in Figure 3. We determined V = 15.92 km/s at T = 0.95 eV and P = 397.39 GPa, and these values agreed with the adiabatic sound velocity V = 16.22 km/s extracted from the fitting dispersion relation, shown in Figure 3. The adiabatic sound velocity increaseed at the elevated temperature and density. The adiabatic sound velocity rangeed from 15.92 km/s to 25.22 km/s by EOS and rangeed from 16.22 km/s to 23.43 km/s by fitting dispersion relation. The adiabatic sound velocity difference between the two methods was within 8%. The adiabatic sound velocity by ion-ion dynamic structure factors may be limited by the size of the supercell (which limited the numbers of small q) and time steps, and even data post-process.

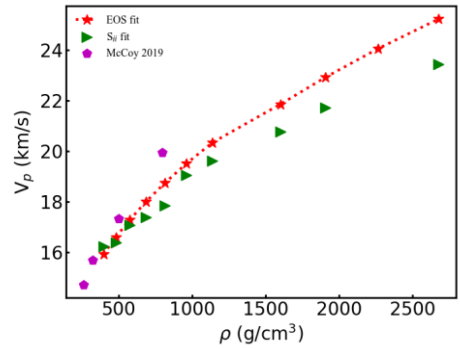

Figure 3 Sound velocity of beryllium along the Hugoniot. The 'EOS' and '$S_{ii}$' labels correspond to adiabatic sound velocity calculated by the EOS and ion-ion dynamics structure factor, respectively. The 'McCoy 2019' label is experimental data from [1].

3.4. Optical properties

The Hugoniot state corresponded to the first stage of hypervelocity impact, which corresponds to a warm dense plasma and was constructed to continuous optical spectrums. The ab initio calculated absorption spectrums of beryllium along the Hugoniot were continuous without any line characteristic, which has been reported [8]. The beryllium atoms in these shocked warm dense states were closely condensed as the same properties of black body radiation, which show a continuous characteristic. At the later stage of the hypervelocity impact, the plasmas were quickly expanded at high temperature and low density, the plasmas show an atomic or ionic characteristic, which corresponded to a line characteristic spectrum. The line spectrum of the hypervelocity impact was constructed on the later stage of the impact which corresponds to a low density gas and can be calculated by the SpectraPlot [25]. The atomic optical spectroscopy calculated by SpectraPlot [25] was different from our results about beryllium along the Hugoniot state. The energy-loss spectrums were also collected, which also show a notable peak around 30 eV, and high temperature/pressure states always shifted to the high energy.

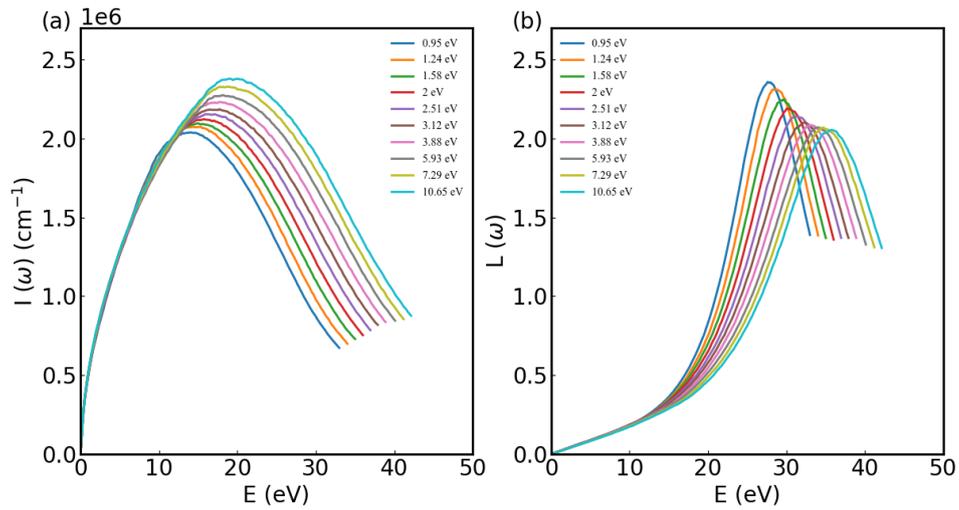

Figure 4 Ab initio calculated (a) absorption coefficient and (b) energy loss spectrum of beryllium along the Hugoniot state.

4. Conclusions

In conclusion, we have determined the static structure factors, ion-ion dynamic structure factors, adiabatic sound velocity, and optical properties of beryllium along shocked Hugoniot by ab initio molecular simulations. The static structure factors showed that beryllium atoms are randomly distributed in the cell. The calculation results of structure factors can be used for a direct comparison for the inelastic x-ray scattering spectra experiments. The adiabatic sound velocity calculated from dispersion relation for the ion-ion dynamic structure factors and EOS was equivalent. From the equivalence of calculation method about adiabatic sound velocity, the dynamic structure factors method is appropriate when the adiabatic sound velocity at a specified state is needed, and the EOS method is appropriate when the adiabatic sound velocity at a certain range is needed. The optical spectrums were continuous along the Hugoniot without any line characteristics, which can provide data for the first stage of hypervelocity impact. However, when the line atomic optical spectrum was needed, more calculations should be conducted.

**Acknowledgement**

This work was support by the National Natural Science Foundation of China (NSFC) [grant numbers 11975058, 11775031 and 11625415], and the fund of Key Laboratory of Computational Physics [grant number 6142A05RW202103]. We thanked for the fund support from Laboratory of Computational Physics in Institute of Applied Physics and Computational Mathematics.